\documentclass[letterpaper,11pt]{article}
\pdfoutput=1
\usepackage{jhepmod} 

\usepackage{booktabs}
\usepackage[english]{babel}
\usepackage{amsmath,amssymb,amsbsy,amstext, amsthm, simplewick, amsfonts}
\usepackage{hyperref}
\usepackage{graphicx}
\usepackage{subcaption}
\usepackage{siunitx}
\usepackage{upgreek}
\usepackage{framed}
\usepackage{wrapfig}
\usepackage{multirow}
\usepackage{bbm}
\usepackage[utf8x]{inputenc}
\usepackage{selinput}
\usepackage{bm}
\usepackage{float}
\usepackage{yfonts}
\usepackage{mathtools}
\usepackage{enumitem}   
\usepackage{longtable}
\usepackage{tabularx}
\usepackage{afterpage}

\usepackage{comment}
\usepackage{soul}

\def\k{{\boldsymbol k}}
\def\q{{\boldsymbol q}}

\def\cO{{\cal O}}

\def\beq{\begin{equation}}
\def\eeq{\end{equation}}
\def\be{\begin{equation}}
\def\ee{\end{equation}}

\def\be{\begin{equation}}
\def\ee{\end{equation}}
\def\ben{\begin{eqnarray}}
\def\een{\end{eqnarray}}

\def\oh{\bf \hat\Omega}

\def\2p{{(2\pi)^2}}

\def\be{\begin{equation}}
\def\ee{\end{equation}}
\def\beq{\begin{equation}}
\def\eeq{\end{equation}}
\def\ben{\begin{eqnarray}}
\def\een{\end{eqnarray}}
\def\bes{\begin{subequations}}
\def\ees{\end{subequations}}

\def\oh{{\hat\Omega}}
\def\nn{{\nonumber}}

\newcommand{\beqa}{\begin{eqnarray}}
\newcommand{\eeqa}{\end{eqnarray}}

\def\ikap0{{\cal J}_{\theta_0}(r)}

\def\one1{\langle \kappa_{(i)}\kappa_{(j)} \rangle}
\def\one{{[\bar \xi^{(ij)}]}}

\def\ba{\begin{eqnarray}}
\def\ea{\end{eqnarray}}

\def\2p{{(2\pi)^2}}

\def\be{\begin{equation}}
\def\ee{\end{equation}}

\def\beq{\begin{equation}}
\def\eeq{\end{equation}}

\def\ben{\begin{eqnarray}}
\def\een{\end{eqnarray}}

\def\oh{{\hat\Omega}}
\def\nn{{\nonumber}}

\def\2p{{(2\pi)^2}}

\def\kp{{k_{\perp}}}

\def\kp{{\kappa}}

\def\nn{\nonumber\\}

\usepackage{xcolor}
\usepackage{xcolor}

\title{Weak Lensing Trispectrum and Kurt-Spectra}
\author{Dipak Munshi$^{a,1}$, Hayden Lee$^{b,c,2}$, Cora Dvorkin$^{b,3}$, Jason D. McEwen$^{a,d,4}$} 
\affiliation{$^{a}$Mullard Space Science Laboratory, University College London, \\
  Holmbury St Mary, Dorking, Surrey RH5 6NT, UK}
\affiliation{$^{b}$Department of Physics, Harvard University, 17 Oxford Street,\\
  Cambridge, MA 02138, USA}
  \affiliation{$^{c}$Kavli Institute for Cosmological Physics, University of Chicago, Chicago, IL 60637, USA}
\affiliation{$^{d}$The Alan Turing Institute, Euston Road, London NW1 2DB, UK}

\emailAdd{$^1$d.munshi@ucl.ac.uk,
  $^2$haydenl@uchicago.edu, $^3$cdvorkin@g.harvard.edu,
  $^4$jason.mcewen@ucl.ac.uk}

\abstract{We introduce two kurt-spectra to probe fourth-order statistics of weak lensing convergence maps.
  Using state-of-the-art numerical simulations, we study the shapes of these kurt-spectra as a function of source redshifts and smoothing angular scales. We employ a pseudo-$C_{\ell}$ approach to estimate the spectra from realistic convergence maps in the presence of an observational mask and noise for stage-IV large-scale structure surveys. We compare these results against theoretical predictions calculated using the FFTLog formalism, and find that a simple nonlinear clustering model---the hierarchical ansatz---can reproduce the numerical trends for the kurt-spectra in the nonlinear regime. 
  In addition, we provide estimators for beyond fourth-order spectra where no definitive analytical results are available, and present corresponding results from numerical simulations.
  }

\begin{document}

\maketitle
\flushbottom
\newpage
\section{Introduction}
\label{sec:into}

Despite the huge amount of progress in cosmology in the past few decades, there still remain many outstanding questions. These include the nature of dark matter (DM), the source of the accelerated expansion of the universe, and the physics of the early universe.
In addition, the sum of the neutrino masses~\cite{nu} remains unknown. It is expected that the operational weak lensing surveys, including the {Subaru Hyper Suprime-Cam Survey}\footnote{\href{http://www.naoj.org/Projects/HSC/index.html}{\tt http://www.naoj.org/Projects/HSC/index.html}}~(HSC)~\cite{HSC},
the Dark Energy Survey\footnote{\href{https://www.darkenergysurvey.org}{\tt https://www.darkenergysurvey.org}}~(DES)~\cite{DES},
the Dark Energy Spectroscopic Instrument~(DESI)\footnote{\href{http://desi.lbl.gov}{\tt http://desi.lbl.gov}}~\cite{DESI:2016fyo},
the Prime Focus Spectrograph\footnote{\href{http://pfs.ipmu.jp}{\tt http://pfs.ipmu.jp}}~\cite{Tamura:2016wsg},
the Kilo-Degree Survey (KiDS)~\cite{KIDS}, as well as near-future Stage-IV large-scale structure (LSS)
surveys such as \textit{Euclid}\footnote{\href{http://sci.esa.int/euclid}{\tt http://sci.esa.int/euclid}}~\cite{Euclid},
the Vera C.~Rubin Observatory\footnote{\href{http://www.lsst.org/llst home.shtml}{\tt {http://www.lsst.org/llst home.shtml}}}~\cite{LSSTTyson},
and the Roman Space Telescope\footnote{\href{https://roman.gsfc.nasa.gov}{\tt https://roman.gsfc.nasa.gov}}~\cite{WFIRST,Spergel}, will improve our understanding to many of the questions that cosmology is facing from high-precision measurements of the intervening mass distribution of the universe.

Weak lensing observations target the low-redshift 
universe and small scales, where density perturbations are mostly in the nonlinear regime and
the statistics are highly non-Gaussian~\cite{Munshi_Review}. Hence, unlike the
high-redshift cosmic microwave background (CMB) radiation, the power spectrum
alone is not sufficient to distill the entire information content of the data. For this reason, many different estimators have been developed to probe higher-order statistics of weak lensing maps~\cite{Carbone}. Initial work in this
direction focused primarily on analyzing various statistics that are directly related to the bispectrum such as the
integrated bispectrum, a skew-spectrum estimator, and a morphological estimator.

With the increase in high-quality data from ongoing surveys, it is now becoming possible to probe statistics beyond the bispectrum, e.g.\ the trispectrum, which represents the connected contribution to the four-point correlation function in the Fourier (or harmonic) domain~\cite{Hu_Trispectrum, HuOkamoto}. However, detection and characterization of individual trispectral modes (represented by a quadrilateral)
remain computationally challenging. To this end, compressed statistics such as the generalization of the skew-spectrum to fourth order---known also as the {\it kurt-spectra}---were introduced in the context of 21-cm surveys~\cite{Cooray1,Cooray2}. Two types of such spectra were implemented in~\cite{myTri1, myTri2} and have already been studied in the context of primordial non-Gaussianity (PNG), where the main motivation was to put independent constraints on the two shapes of local non-Gaussianity parameterized by $\tau_{\rm NL}$ and $g_{\rm NL}$~\cite{Planck13a, Planck:2015zfm, Planck:2019kim, Smith:2015uia}. 
The situation for the gravity-induced trispectrum is very similar in modified gravity theories, where more than one parameter characterizes the gravity-induced trispectrum. The use of the two kurt-spectra can lift the degeneracy and provide an important consistency check for the constraints from lower-point statistics.

Moreover, the kurt-spectra were used to detect the lensing-induced secondary \linebreak non-Gaussianity; for example, their application to the WMAP 7-year temperature maps resulted in the first direct constraints of the CMB lensing potential power spectrum~\cite{Joseph_Lensing}.
Planck used a similar technique for their analysis~\cite{Planck13b}, and the corresponding fourth-order real-space correlation functions were used in the context of CMB secondaries~\cite{Spergel} to separate the lensing of the CMB from the Ostriker-Vishniac effect. In real space, these correlation functions were also studied to arbitrary order in the context of gravity-induced non-Gaussianity in the LSS using the standard perturbation theory (SPT)~\cite{Bernardeau_Bias, MMC},
but in the limit of large angular separations~\cite{MunshiMcEwen}. Theoretical modeling of trispectra has attracted more attention recently~\cite{LeeDvorkin, Verde}.
Beyond the SPT, the effective field theory (EFT) based approach has been used to model the trispectrum~\cite{EFTtri, Baldauf_Tri}, as well as the halo-model~\cite{halo} and the hierarchical ansatz based approaches~\cite{hier}, which are valid in the quasi-linear and (highly) nonlinear regimes.

In addition to the summary statistics listed above, in recent years a number of novel modeling techniques have gained popularity. These include Bayesian hierarchical modeling, likelihood-free or forward modeling approaches \cite{likefree1,likefree2,Porqueres:2021clw,Taylor:2019mgj,DiazRivero:2020oai}, as well as wavelet phase harmonics \cite{PhysRevD.102.103506} and
the scattering transform  \cite{mallat2012group,Cheng:2020qbx, ChengChengSiHao:2021hja} (see also \cite{saydjari2021classification,Valogiannis:2021chp,Regaldo-SaintBlancard:2020dlb,Allys:2019fna} for other works applying scattering transform-type statistics to different astrophysical observables).\footnote{Many studies have also focused on one-point statistics for probing higher-order statistics.
These include the well-known real-space one-point statistics such as the cumulants~\cite{Barber1} or two-point cumulant correlators as well as the associated probability distribution function~\cite{Uhlemann1}, the peak-count statistics~\cite{peak_count}, and morphological estimators~\cite{morph}.}

One of the primary aims of this paper is to generalize the kurt-spectra used in~\cite{myTri1, myTri2} in the presence of a realistic {\em Euclid}-type mask and noise.
Using a suite of state-of-the-art numerical simulations, we study the gravity-induced non-Gaussianity using weak lensing convergence $\kappa$ maps. 
The gravity-induced signal is sufficient to saturate the Fisher bounds for all-sky low-noise maps expected from {\em Euclid}. Note that this might not be the case for ongoing surveys that are noise dominated
and cover a small fraction of the sky; for these studies, an optimization in line with what was presented in~\cite{myTri1, myTri2} may be necessary.
Such procedures are, however, only optimal in the limit of weak non-Gaussianity, and may not be relevant for signal-dominated data from future surveys. 
We will thus simply stick to sub-optimal estimators in this work, and further neglect PNG though it can be incorporated the same framework. 
Generalizations of our estimators to spectroscopic galaxy redshift surveys such as BOSS\footnote{\href{http://www.sdss3.org/surveys/boss.php}{\tt http://www.sdss3.org/surveys/boss.php}}~\cite{SDSSIII}
or WiggleZ\footnote{\href{http://wigglez.swin.edu.au/}{\tt http://wigglez.swin.edu.au/}}~\cite{WiggleZ} that probe the mass distribution of galaxies as biased tracers~\cite{biasreview} are left for future work.

This paper is organized as follows. In~\textsection\ref{sec:matter}, we review the formalism for computing angular trispectra and describe the analytical modeling of gravity-induced trispectra that we adopt in this work. In~\textsection\ref{sec:kurt}, we describe the trispectrum of weak lensing convergence and introduce the kurt-spectra.
The ray-tracing simulations that we have used and the results obtained are presented in~\textsection\ref{sec:disc}. We conclude and discuss future prospects in~\textsection\ref{sec:conclusions}.

\section{Formalism}
\label{sec:matter}

We begin with a short review of $n$-point correlation functions in harmonic space in \S\ref{sec:harmonic}, focusing on the case $n=4$. We describe the computational methods for the angular trispectrum in \S\ref{sec:comp} and the theoretical models for the matter trispectrum used in our analysis in \S\ref{sec:model}.

\subsection{Correlators in Harmonic Space}\label{sec:harmonic}

In cosmological observations, a projected observable $\cal O$ located at some redshift $z$ is measured as a function of its angular position on the celestial sphere. This is usually thought as tracing the underlying matter density contrast $\delta$ integrated along the line-of-sight direction $\hat n$, weighted by some kernel $W_{\cal O}$ as 
\begin{align}
    \cO(z,\hat n) = \int_0^{\chi(z)} d\chi'\, W_{\cO}(\chi') \delta(\chi',\chi'\hat n)\, ,
\end{align}
where $\chi$ is the comoving radial distance. It is often useful to take advantage of the spatial isotropy of the celestial sphere and work in harmonic space, which allows for a spectral analysis. Expanding the real-space observable in spherical harmonics as $\cO(z,\hat n)=\sum_{\ell=0}^\infty\sum_{m=-\ell}^\ell\cO_{\ell m}^{(z)} Y_{\ell m}^{\phantom{z}}(\hat n)$, the harmonic coefficients are obtained through the projection integral
\begin{align}
    \cO_{\ell m}^{(z)}=4\pi i^\ell \int_0^{\chi(z)} d\chi'\, W_\cO (\chi') \int\frac{d^3k}{(2\pi)^3}\,j_\ell(k\chi')Y_{\ell m}^*(\hat k)\delta(\chi',\k)\, ,
\end{align}
where we have Fourier transformed $\delta$ and projected the plane waves onto the spherical harmonics basis. 
The $n$-point function in harmonic space is then obtained by taking the expectation value of a product of harmonic coefficients as
\begin{align}\label{npt}
    \langle\cO_{\ell_1m_1}^{(z_1)}\cdots\cO_{\ell_n m_n}^{(z_n)}\rangle = (4\pi)^ni^{\ell_{1\cdots n}}\int\prod_{i=1}^n\left[\frac{d\chi_i'd^3k_i}{(2\pi)^3}W_{\cO_i}(\chi_i')j_{\ell_i}(k_i\chi_i')Y_{\ell_im_i}^*(\hat k_i)\right]\langle\delta_1\cdots\delta_n \rangle\, ,
\end{align}
where we defined $\ell_{1\cdots n}\equiv\ell_1+\cdots+\ell_n$, $\cO_i \equiv \cO_{\ell_i m_i}^{(z_i)}$, and $\delta_i \equiv \delta(z_i,\k_i)$. In general, it is a challenging task to evaluate the convoluted multi-dimensional projection integrals appearing in \eqref{npt} for $n\ge 3$. However, as we will shortly review, these integrals become factorized for {$\delta$-correlation} functions that respect a certain separability condition, rendering the computation of higher-point functions much more feasible.

Our primary interest in this work will be the case $n=4$, namely the angular trispectrum in harmonic space. A nice feature of a harmonic-space analysis is that we can completely factor out the azimuthal dependence and write the harmonic-space trispectrum as~\cite{Hu_Trispectrum, Mitsou:2019ocs}
\begin{align}
\langle \cO_{\ell_1m_1}^{(z_1)} \cdots \cO_{\ell_4 m_4}^{(z_4)}\rangle
&= \sum_{LM} (-1)^M  \left ( \begin{array}{ c c c }
     \ell_1 & \ell_2 & L \\
     m_1 & m_2 & M
\end{array} \right )
\left ( \begin{array}{ c c c }
     \ell_3 & \ell_4 & L \\
    m_3 & m_4 & -M
\end{array} \right ) T^{\ell_1\ell_2}_{\ell_3\ell_4}(L) \label{tridecomp} \\
& = \sum_{LM} (-1)^M  \left ( \begin{array}{ c c c }
     \ell_1 & \ell_2 & L \\
     m_1 & m_2 & M
\end{array} \right )
\left ( \begin{array}{ c c c }
     \ell_3 & \ell_4 & L \\
    m_3 & m_4 & -M
\end{array} \right ) P^{\ell_1\ell_2}_{\ell_3\ell_4}(L)
+ (2 \leftrightarrow 3) + (2 \leftrightarrow 4)\, ,\nonumber
\end{align}
where $\sum_{LM}\equiv \sum_{L=0}^\infty\sum_{m=-L}^L$ and the rounded-bracketed matrices represent the Wigner 3-j symbols. In the second line, we have split the trispectrum into three different pairings of multipoles using permutation symmetry. As can be seen from \eqref{npt}, the input for the angular trispectrum is the matter trispectrum in Fourier space, which we can write as
\begin{align}
    \langle \delta_1\cdots\delta_4 \rangle =  T_{\delta}(\{z_i,\k_i\}) \times (2\pi)^3 \delta_D(\k_1+\cdots+\k_4)\, ,
\end{align}
where $\{z_i,\k_i\}=\{z_1,k_1,\cdots,z_4,k_4\}$ denotes the set of arguments and $\delta_D$ represents the Dirac delta function that ensures spatial translational invariance. In accordance with \eqref{tridecomp}, the trispectrum can be decomposed into three different channels as
\begin{align}
   T_{\delta}(\{z_i,\k_i\})= T^{(s)}_{\delta}(\{z_i,\k_i\}) + T^{(t)}_{\delta}(\{z_i,\k_i\})
  + T^{(u)}_{\delta}(\{z_i,\k_i\})\, ,\label{Tdecomp}
\end{align}
where we defined $s=|\k_1+\k_2|$, $t=|\k_1+\k_4|$, $u=|\k_1+\k_3|$, and different channels are related by the permutations $2\leftrightarrow 3$ and $2\leftrightarrow 4$. These $s,t,u$-channel contributions are thus in one-to-one correspondence with $P^{\ell_1\ell_2}_{\ell_3\ell_4}(L)$, $P^{\ell_1\ell_4}_{\ell_3\ell_2}(L)$, $P^{\ell_1\ell_3}_{\ell_2\ell_4}(L)$, respectively, in \eqref{tridecomp}.

Note that the (reduced) angular trispectrum $T^{\ell_1\ell_2}_{\ell_3\ell_4}(L)$ in \eqref{tridecomp} is defined with respect to a particular pairing of multipoles that corresponds to the $s$-channel. 
Unlike in Fourier space, different multipole pairings in harmonic space do not contribute to the total trispectrum in a simple additive manner, but instead they are related by~\cite{Hu_Trispectrum}
\begin{align}
\label{Tdecomp} &T^{\ell_1\ell_2}_{\ell_3\ell_4}(L) =  P^{\ell_1\ell_2}_{\ell_3\ell_4}(L)  \\
    &\quad + (2L+1) \sum_{L'} \left( (-1)^{\ell_2+\ell_3} \left \{ \begin{array}{ c c c }
     \ell_1 & \ell_2 & L \\
     \ell_4 & \ell_3 & L'
\end{array} \right \} P^{\ell_1\ell_3}_{\ell_2\ell_4}(L')  +
(-1)^{L+L'}   \left \{ \begin{array}{ c c c }
    \ell_1 & \ell_2 & L \\
     \ell_3 & \ell_4 & L'
  \end{array} \right \} P^{\ell_1\ell_4}_{\ell_3\ell_2}(L')\right),\nonumber
\end{align} 
where the curly-bracketed matrices represent the Wigner 6-j symbols.  Despite there being a complicated relation amongst three channels, it turns out that a simplifying approximation
$T^{\ell_1\ell_2}_{\ell_3\ell_4}(L) \approx  P^{\ell_1\ell_2}_{\ell_3\ell_4}(L)$ is often adequate for most purposes (see e.g.~\cite{Hu_Trispectrum, Joseph_Lensing}). This dramatically simplifies numerical analyses of harmonic-space trispectra; we adopt this approximation in this work.

As was alluded to above, the projection integrals in \eqref{npt} become dramatically simplified for $\delta$-correlation functions that are {\it separable}. Roughly speaking, a separable correlation function means that its individual terms can be expressed as a product of some functions of momentum variables. The precise separability condition for the matter trispectrum put forward in~\cite{LeeDvorkin} is that individual terms, say, in the $s$-channel can be expressed as
\ben
T_{\delta}^{(s)}(\{z_i,\k_i\})\, \supset \,  f_1(z_1,k_1) \cdots f_4(z_4,k_4) f_s(s)t^{2J}\, ,\label{separable}
\een
where $J$ is a non-negative integer for local interactions.\footnote{See e.g.~\cite{Regan:2010cn, Smith:2015uia} for alternative separability criteria for trispectra.} The gravitationally-induced matter trispectrum that is relevant for our purpose has $J=0$. In this case, the angular and radial integrals in \eqref{npt} become completely factorized, with the separable term \eqref{separable} in Fourier space resulting in the following form in harmonic space~\cite{LeeDvorkin}:
\ben
P^{\ell_1\ell_2}_{\ell_3\ell_4}(L)
= \frac{ g^{\ell_1\ell_2L} g^{\ell_3\ell_4L}}{(2\pi^2)^5}
\int_0^\infty dr\, r^2 I_{\ell_1}^{(1)}(r) I_{\ell_2}^{(2)}(r) 
\int_0^\infty dr'\, r'^2 I_{\ell_3}^{(3)}(r') I_{\ell_4}^{(4)}(r')J_L^{(s)}(r,r') \, ,\label{separableP}
\een
where
 \begin{align}
g^{\ell_1\ell_2\ell_3}= \sqrt{\frac{(2 \ell_1 +1)(2 \ell_2 +1)(2 \ell_3 +1) }{4 \pi}}
\left ( \begin{array}{ c c c }
     \ell_1 & \ell_2 & \ell_3 \\
     0 & 0 & 0
  \end{array} \right)
  \end{align}
is a geometric factor and
 \begin{align}
 I^{(i)}_\ell(r) &= 4\pi  \int_0^{\chi(z)} d\chi' \, W_{\cO}(\chi')
\int_0^{\infty} dk\, k^2 f_i(z(\chi'),k) j_\ell(k\chi') j_{\ell}(kr)\, , \label{Iint}\\
J^{(s)}_L(r,r') &= 4\pi \int_0^{\infty} ds \, s^2  f_s(s) j_L(sr) j_L(sr')\, ,\label{Jint}
\end{align}
are the projection radial integrals, with $j_\ell$ the spherical Bessel function. 
Typically, the functions of momenta take the form $f_i(z,k)=D_+^{m_i}(z)k^{2n_i}P_\delta^{p_i}(k)$, where $D_+$ is the linear growth function, $P_\delta$ is the matter power spectrum, and $n_i,m_i,p_i$ are integers.\footnote{More generally, one needs to introduce a scale-dependence in the linear growth function due to e.g.~massive neutrinos. Such cases can be efficiently dealt with by the use of a polynomial approximation that separates the scale and redshift dependences; see~\cite{Levi:2016tlf, Chen:2021vba} for details.} This allows us to further simplify the double integral in \eqref{Iint}, as we show next.

A few comments about Eq.(\ref{Tdecomp}) and the subsequent approximation are in order.
  The evaluation of the expression given in Eq.(\ref{Tdecomp}) for the trispectrum is 
  computationally challenging. This approximation was introduced in the context of CMB studies.
  In the case of low-redshift weak lensing studies, the situation is even more difficult
  due to the line-of-sight integration.

An alternative to this approach was introduced in
  \citep{Matsubara} (also see \citep{Munshi_CMB_Lensing}), where the
  spherical sky expression is replaced by the corresponding flat-sky approximations, and the 3j- and 6j-symbols are replaced by Dirac delta functions.
 The summations that appear in all-sky calculations are subsequently replaced by integrals that can be carried out using higher-dimensional Monte Carlo
 computations.

 The primary aim of this article is to introduce the kurt-spectra and their higher-order analogs.
   A more accurate modeling will be taken up in the future.
   This is likely to take a hybrid approach, where the entire range of $\ell$ values is split into a low-$\ell$ and a high-$\ell$ regime. The low-$\ell$ ($\ell<100$) regime can be tackled using
   an all-sky calculation, where 6j-symbols computations are feasible, and
   a flat-sky method will be used for the high-$\ell$ regime.

   
\subsection{Computational Methods}\label{sec:comp}

As we just saw, the computation of harmonic-space trispectra amounts to evaluating the projection integrals of the form \eqref{Iint} and \eqref{Jint}. Naively, the presence of highly oscillatory Bessel functions in the integrand makes a direct numerical integration quite difficult, especially for high multipoles. Over recent years, efficient semi-analytic methods for evaluating these projection integrals have been developed in~\cite{Assassi:2017lea, GrasshornGebhardt:2017tbv, Schoneberg:2018fis, LeeDvorkin} based on the algorithm known as the FFTLog~\cite{Hamilton:1999uv}, with the goal of computing angular observables in cosmology in a numerically fast and accurate way (see \cite{Leistedt:2012zx, Campagne:2017xps, Slepian:2018vds, DiDio:2018unb, Fang:2019xat, Deshpande:2020jjs, GrasshornGebhardt:2020wsw, Fang:2020vhc, Montanari:2020uez, Umeh:2020zhp, Chen:2021vba} for related developments and applications of these methods). 

The basic idea of these methods is to discrete Fourier transform (in $\log k$) the matter power spectrum over some finite interval $[k_{\rm min},k_{\rm max}]$ as
\begin{align}
    P_\delta(z,k)  &\approx \sum_{m=-N/2}^{N/2} c_m(z)k^{-b+i\eta_m}\, , \quad \eta_m\equiv \frac{2\pi m}{\log(k_{\rm max},k_{\rm min})}\, ,
\end{align}
where $b$ is a real parameter introduced for convenience and the coefficients of the transform are given by
\begin{align}
    c_m(z) & = \frac{2-\delta_{|m|,N/2}}{2N}\sum_{n=0}^{N-1}P_\delta(z,k_n)k_n^bk_{\rm min}^{-i\eta_m}e^{-2\pi i mn/N}\, .
\end{align}
Essentially, the FFTLog approximates the matter power spectrum in terms of a finite number of complex power-law functions, with a sub-percent accuracy for $N=O(10^2)$. The usefulness of this approximation is that the momentum integrals in \eqref{Iint} and \eqref{Jint} can now be done analytically for each complex power-law function, allowing us to express them as
\begin{align}
    I_\ell^{(i)}(r) &\approx \sum_m c_m\int_0^{\chi(z)} d\chi'\, \chi'^{-\nu_m} D_+^{m_i}(z(\chi'))W_\cO(\chi'){\sf I}_\ell(\nu_m,\tfrac{\chi'}{r})\, ,\\
    J_L(r,r') &\approx \sum_m c_m r^{-\nu_m}{\sf I}_L(\nu_n,\tfrac{r'}{r})\, ,
\end{align}
with~\cite{Assassi:2017lea}
\begin{align}
    {\sf I}_\ell(\nu,w) &\equiv 4\pi \int_0^\infty d x\, x^{\nu-1}j_\ell(x)j_\ell(wx)\nn[3pt]
&=\frac{2^{\nu-1}\pi^2\Gamma(\ell+\frac{\nu}{2})}{\Gamma(\frac{3-\nu}{2})\Gamma(\ell+\frac{3}{2})} w^\ell\, {}_2F_1(\tfrac{\nu-1}{2},\ell+\tfrac{\nu}{2},\ell+\tfrac{3}{2},w^2)\, ,
\end{align}
for $w\le 1$, and ${}_2F_1$ denotes the Gauss hypergeometric function. For $w>1$, one uses the property ${\sf I}_\ell(\nu,w) = w^{-\nu}{\sf I}_\ell(\nu,\frac{1}{w})$. Since the hypergeometric function is a smooth function whose analytic properties are well known, this provides an efficient way to compute the projection integrals, avoiding the need to directly integrate the Bessel functions. 

If we restrict to large multipoles and sufficiently smooth line-of-sight kernels, then there is a more widely used approximation to deal with the projection integrals known as the Limber approximation~\cite{Limber:1954zz, LoVerde:2008re}. This amounts to replacing the spherical Bessel functions $j_\ell$ in the integrands with Dirac delta functions, $j_\ell(x) \approx \sqrt{\pi \over 2\ell}\delta_D(\ell-x)$, which leads to
\begin{align}
 I_{\ell}^{(i)}(r) &\approx \frac{2\pi^2}{r^2} D_{+}^{m_i}(r) W_{\cO}(r) \tilde f_i(\ell/ r)\, , \\
J_{L}^{(s)}(r,r') &\approx \frac{2\pi^2}{r^2}  f_s(L/r)  \delta_D(r-r')\, ,
\end{align}
where $\tilde f_i$ is $f_i$ evaluated at $z=0$. Note that what used to be the $\chi$ integrand does not carry any multipole dependence, allowing us to factor it out from individual terms. Moreover, the leftover delta function in $J_L^{(s)}$ removes one of the radial integrals in \eqref{separableP}. As a consequence,
the Limber-approximated angular trispectrum simply reduces to the following one-dimensional integral (at tree level):
\ben
P^{\ell_1\ell_2}_{\ell_3\ell_4}(L)  \approx g^{\ell_1\ell_2 L} g^{\ell_1\ell_2 L}
\int_0^{\infty} \frac{dr}{r^6}\, D^6_{+}(r) W^4_{\cO}(r)
{\tilde T}^{(s)}_{\delta}(\tfrac{\ell_1}{r}, \cdots, \tfrac{\ell_4}{r},\tfrac{L}{r}) \, ,\label{Plimber}
\een
where ${\tilde T}^{(s)}_{\delta}$ is defined to be the purely momentum-dependent part of the matter trispectrum
\begin{align}
    {\tilde T}^{(s)}_{\delta}(k_1,\cdots,k_4,s)=T^{(s)}_{\delta}(\{z_i,\k_i\})|_{z_i=0}\, ,
\end{align}
with the normalization $D_+(0)=1$. Similar expressions exist for the $t$- and $u$-channels. We will show a comparison of the FFTLog and Limber approximation in \S\ref{sec:tri}.

\subsection{Theoretical Model for Matter Trispectrum}\label{sec:model}

Cosmological angular trispectra are obtained by projecting the matter trispectrum along the line of sight. While this in principle involves integrating over all momenta, typical scales are related by $\ell\sim k\chi(z)$ between the harmonic and Fourier domains.
Depending on the harmonics, two distinct theoretical models for matter clustering are then relevant in the quasi-linear and nonlinear regimes. Let us briefly review the models that we consider in our study.

\paragraph{Quasi-linear regime}
At sufficiently large scales, cold dark matter behaves as an effective pressureless fluid, and its gravitational evolution is governed by the Newtonian fluid equations of motion. In the standard perturbation theory (SPT) framework~\cite{bernardeaureview}, these equations are solved perturbatively by expanding the nonlinear density contrast in terms of the linear solution $\delta^{(1)}$ as
\begin{align}
    \delta(z,\k) =\sum_{n=1}^\infty D_+^n(z) \int_{\q_1,\cdots,\q_n}(2\pi)^3\delta_D(\k-\q_{1\cdots n})F_n^{\rm sym}(\q_1,\cdots,\q_n)\delta^{(1)}(\q_1)\cdots \delta^{(1)}(\q_n)\, ,
\end{align}
where $\int_{\q_1,\cdots,\q_n}\equiv\int\prod_{i=1}^n\frac{d^3q_i}{(2\pi)^3}$, $q_{1\cdots n}\equiv\q_1+\cdots\q_n$ and $F_n^{\rm sym}$ is the symmetrized SPT kernel~\cite{Goroff:1986ep, Jain:1993jh}. The Einstein-de Sitter (matter domination) approximation is typically used so that the temporal and spatial dependences become factorized in the way above.
At tree level, there are two contributions to the matter trispectrum
\ben
T_{\delta}^{\rm SPT}(\{z_i,\k_i\}) =   T_{3111}(\{z_i,\k_i\})+T_{2211}(\{z_i,\k_i\})\, ,
\een
which follows from the two distinct ways of expanding $\delta$. 
We quote here the standard results from the SPT~\cite{bernardeaureview}
\begin{align}
  T_{3111}(\{z_i,\k_i\}) &= 6 D_1 D_2 D_3 D_4^3 P_1 P_2 P_3 F_3^{\rm sym}(\k_1,\k_2,\k_3)  + \text{3 perms}\, ,\\
 T_{2211}(\{z_i,\k_i\}) &= 4 D_1 D_2^2 D_3 D_4^2P_1P_3P_sF_2^{\rm sym}(\k_1, -\k_{12})F_2^{\rm sym}(\k_3, \k_{12}) + \text{11 perms}\, ,
\end{align}
where we used the notation $D_{i}\equiv D_+(z_i)$ and $P_{i} \equiv P_{\delta}^{\rm L}(k_i)$ for the linear matter power spectrum. 
To utilize the separability, we express the above trispectrum in terms of momentum magnitudes and write the total tree-level trispectrum as
\begin{align}
    T_\delta^{\rm SPT} (\{z_i,\k_i\}) &=   \Big[T_{\rm exchange}^{(s)}(k_1,k_2,k_3,k_4,s) + \text{2 perms}\Big]+ T_{\rm contact}(k_1,k_2,k_3,k_4)\, ,
\end{align}
where the labels ``exchange'' and ``contact'' refer to the parts that depend on the internal momenta and that does not, respectively. Explicit expressions of the matter trispectrum in these variables can be found in~\cite{LeeDvorkin}.

\paragraph{Nonlinear regime}

In the nonlinear regime $\delta\gtrsim 1$, the perturbation theory breaks down and we typically have to resort to phenomenological models or fitting functions. 
Unlike for the bispectrum, there currently exists no precise fitting function available for the matter trispectrum that smoothly interpolates between linear and nonlinear scales for all momentum configurations.\footnote{The second-order SPT kernel $F_2$ that characterizes the matter bispectrum has 3 independent tensor structures. A fitting formula is then constructed by endowing each of these structures with a general function momenta that interpolates the linear and nonlinear regimes, see e.g.~\cite{Gil-Marin:2011jtv}. On the other hand, the matter trispectrum depends on the third-order kernel $F_3$ that has 12 independent tensor structures with 6 momentum degrees of freedom, which complicates such a fitting procedure (see \cite{Gualdi:2021yvq} for recent progress on this).} Instead, a nonlinear clustering model known as the {\it hierarchical ansatz} (HA)~\cite{Fry:1983cj} is often invoked. 

In the HA, higher-order spectra of density contrast are written as a sum of product of two-point functions over all possible topologies with different amplitudes. The matter trispectrum then has two contributions given by \cite{FryPeebles, BerSch}
\begin{align}
T_{\delta}^{\rm HA}(\k_1,\cdots,\k_4) & \equiv
R_a\big([P_{\delta}^{\rm NL}(k_1)]^{1+\epsilon} [P_{\delta}^{\rm NL}(k_2)]^{1+\epsilon}
[P_{\delta}^{\rm NL}(k_3)]^{1+\epsilon}+\text{3 perms}\big)\nn[2pt]
&\, + R_b \big([P_{\delta}^{\rm NL}(k_1)]^{1+\epsilon}
[P_{\delta}^{\rm NL}(k_3)]^{1+\epsilon}
[P_{\delta}^{\rm NL}(s)]^{1+\epsilon}+\text{11 perms}\big)\, ,\label{THA}
\end{align}
where $P_{\delta}^{\rm NL}$ denotes the nonlinear matter power spectrum and we have suppressed the redshift dependence.
We will only consider the models with $\epsilon=0$ in this paper. 
The matter trispectrum in the HA is therefore parameterized by two amplitudes $R_a$ and $R_b$, which are assumed to be constant in the strongly nonlinear regime. 
Note that each of the two structures in the HA trispectrum has the same power spectra dependence as $T_{3111}$ and $T_{2211}$ in the tree-level trispectrum, and the amplitudes $R_a$, $R_b$ can be thought as being the angular averages of the SPT kernels~\cite{1992AA, Coles:1999ve, bernardeaureview}, e.g.~$R_a=\langle F_3\rangle_\Omega$, $R_b=\langle F_2\rangle_\Omega^2$, where $\langle F_n\rangle_\Omega\equiv n!\int [\prod_{i=1}^n \frac{d\Omega_i}{4\pi}] F_n(\k_1,\cdots,\k_n)$. 
Sometimes a simpler model is used for which the two amplitude parameters in \eqref{THA} are set equal, $R_a=R_b=Q_4$. It was checked, for instance, in~\cite{Scoccimarro:1999kp} that this choice fits the simulation results well in the nonlinear regime for certain kinematic configurations. For our comparison against simulations, we have taken $Q_4$ as a free parameter to fit the data.

The HA trispectrum has a very different momentum dependence from that of the SPT trispectrum, so that it is only applicable in the strongly nonlinear regime. This is manifest from the soft limit behavior of the two shapes: $T_\delta^{\rm HA} \sim P_\delta(k_1)$ as $k_1\to 0$, while $T_\delta^{\rm SPT} \sim \frac{1}{k_1} P_\delta(k_1)$ in the same limit. In the next section, we compare the shapes of the weak lensing trispectrum arising from these two models.

There are many different HA available in the literature.
  The specific version that we have used was introduced in \cite{Szapudi_Szalay1,Szapudi_Szalay2}.
  In this model the amplitudes associated with diagrams with different
  topologies but with same number of vertices are always equal.
  This is not only assumed at the level of fourth-order (trispectrum) but to an arbitrary order.
  Other models that are well known include the minimal hierarchical model
  which was introduced by \cite{Bernardeau_Schaeffer}. In \citep{MunshiJain1,xMunshiJain2}
  many consequences of minimal hierarchical models for weak lensing were discussed.
Many other versions of HA were tested subsequently. One such model
  assumes $R_b \ne 0$ and $R_a=0$ \cite{Barber_Valgeas1, Barber_Valgeas2}.
 In general, it was found that various choices of $R_a$ and $R_b$ can reproduce the weak lensing statistics
    with varying success. There is no specific HA that can reproduce all observables. Moreover,
    most previous studies concentrate on one-point statistics.
In a related context we have also checked that the kurtosis spectra
  extracted from a simulated log-normal sky is very different
  from the more realistic ray-traced simulations.
  This is important as log-normal simulations
  are routinely used in  field-based inference studies.
  We have also tried the  extension of the log-normal \citep{Skewed_lognormal} model, but they 
  could not be fine-tuned to reproduce the both kurtosis spectra
  for the entire range of redshift and angular harmonics
  probed.
A full modeling of matter trispectrum would involve either
  (a) effective halo model type approach for trispectrum that extends to higher order
  \citep{Philcox}
  or (b) a fitting function for the trispectrum as was done most recently by
   \citep{Takahashi} for the bipectrum. 
For individual shapes of trispectrum \citep{Lewis}, we use estimators
  developed in \citep{Munshi_shape_bispec} for the study of shapes of bispectrum.

 
\section{Weak Lensing Higher-Order Statistics}
\label{sec:kurt}

In this section we compute the weak lensing trispectrum and introduce the kurt-spectra. The weak lensing convergence $\kappa$ is a line-of-sight integration of the underlying density contrast $\delta$, and can be expressed using the lensing kernel $W_{\kappa}$ as
\begin{align}
 \kappa(\hat n) &= \int_0^{\chi_s} d\chi'\, W_{\kappa}(\chi')\delta(\chi', \chi'\hat n)\,, \\
 W_{\kappa}(\chi) &= \frac{3 H_0^2 \Omega_m(1+z(\chi))}{ 2 c^2} \int_\chi^{\chi_s} d\chi' n(\chi')
    \frac{\chi' - \chi}{ \chi'}\, ,
\end{align}
where $n(\chi)$ represents the distribution of lensing sources. In our study, we will assume all sources to be at a single source plane $\chi=\chi_s$, which gives $W_\kappa(\chi)=3H_0^2\Omega_m(1+z(\chi))\chi(\chi_s-\chi)\Theta(\chi_s-\chi)/(2c^2\chi_s)$.
The spherical harmonic coefficients $\kappa_{\ell m}$ of the convergence $\kappa$ map  
is defined through $\kp_{\ell m} \equiv \int d\hat n \, \kp(\hat n) Y_{\ell m}^*(\hat n)$.

\subsection{Weak Lensing Trispectrum}
\label{sec:tri}

\begin{figure}
\begin{center}
  \includegraphics[width=\textwidth]{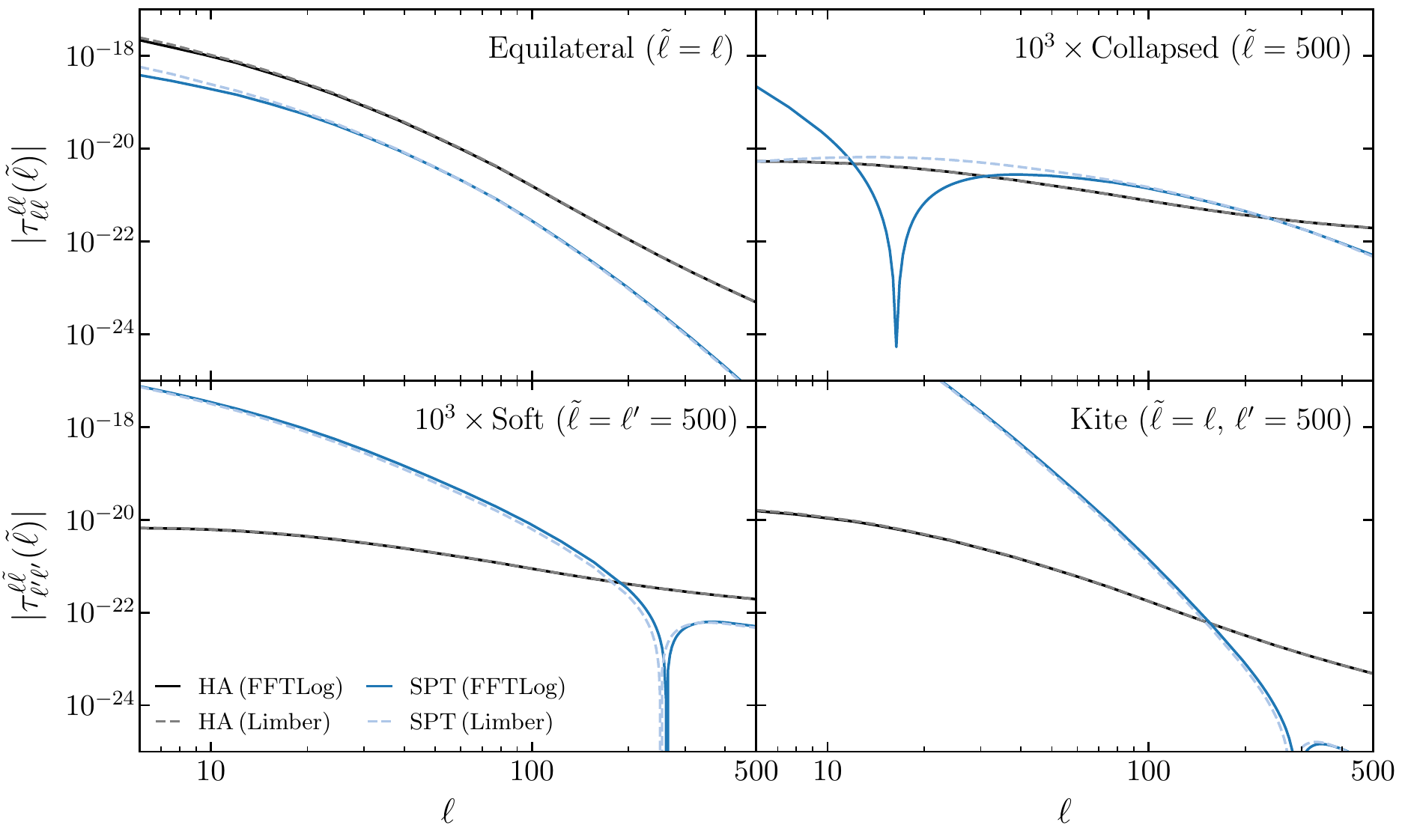}
  \end{center}
  \vspace{-0.25cm}
  \caption{Weak lensing trispectrum at $z_s=1$ after stripping off the geometric factors, c.f.~\eqref{tauell}. Two different models for the matter trispectrum are used: the tree-level result from the standard perturbation theory (SPT) and the hierarchical ansatz (HA) with $R_a=R_b=1$. The solid and dashed lines show the trispectra computed using the FFTLog and the Limber approximation, respectively. 
  }
  \label{fig:tri}
\end{figure}

Using the formalism described in the previous section, we can straightforwardly compute the trispectrum of weak lensing convergence in harmonic space. Figure~\ref{fig:tri} shows the shape of the weak lensing trispectrum for various multipole configurations. We show the reduced part of the trispectrum $\tau^{\ell_1\ell_2}_{\ell_3\ell_4}(L)$ after stripping off the geometric factors, defined by
\begin{align}\label{tauell}
    P^{\ell_1\ell_2}_{\ell_3\ell_4}(L)=g^{\ell_1\ell_2 L}g^{\ell_3\ell_4 L}\tau^{\ell_1\ell_2}_{\ell_3\ell_4}(L)\, ,
\end{align}
at source redshift $z_s=1$. We considered the weak lensing trispectrum arising from two different theoretical models for the matter trispectrum: the tree-level result from the SPT (blue curves) and the HA (black curves) with $R_a=R_b=1$. As the figure shows, these two models lead to drastically different scaling behaviors, clearly highlighting the different domains of applicability of these models. In particular, the HA leads to a less suppressed power compared to the tree-level signal at small scales, as expected for a nonlinear clustering model.

It is interesting to compare the calculations done with the FFTLog and the Limber approximation.\footnote{In Figure~\ref{fig:tri}, we used the FFTLog only for the external multipoles and used the Limber approximation for the internal $L$ to reduce computational costs.} We see that the Limber approximation in general works very well for weak lensing even for very small multipoles, but there is a notable exception: the SPT trispectrum in the collapsed limit $L \ll \ell_1,\cdots,\ell_4$, for which the Limber approximation induces a large deviation. This can be understood from the way the matter trispectrum in Fourier space is projected to harmonic space: the collapsed limit is dominated by terms that scale as inverse powers of $s$ in the SPT trispectrum. It turns out that these terms fully cancel in the rhombus-like configurations $k_1=k_2=k_3=k_4$ in Fourier space, ensuring the infrared safety of the one-loop power spectrum and that the consistency relations are satisfied~\cite{Fujita:2020xtd}. While this continues to be true under the Limber approximation that simply amounts to the substitution $k_i\to\frac{\ell_i}{r}$ (c.f.~\eqref{Plimber}), these terms do not fully cancel in the exact calculation when the projection is taken before taking the equal-multipole limit. The FFTLog method is able to capture this non-cancellation of terms that dominate in the collapsed limit, hence resulting in the large difference between the two computational methods in the collapsed limit. This implies that the Limber approximation of the weak lensing trispectrum from the SPT is highly accurate for most configurations, except in the limit $L\to 0$.\footnote{The same observation was made in \cite{LeeDvorkin}, where it was shown that the Limber approximation fails for the computation of the non-Gaussian covariance of the angular matter power spectrum, which requires evaluating $T^{\ell_1\ell_2}_{\ell_3\ell_4}(L)$ at $L=0$.} For the HA, the Limber approximation was found to be accurate for the multipole configurations we considered.

\subsection{Kurt-Spectra}
\label{sec:kurt2}

While the precise physical content of a correlation function is contained in its entire shape dependence, it is often useful to construct lower-point statistics to estimate higher-point functions. One of the main advantages of such estimators is that they have a much simpler multipole dependence, while still carrying sufficient information to constrain amplitude-like parameters.

The harmonic coefficients of the $n$-th power of the convergence field, $\kappa^n$, can be expressed in terms of $\kappa_{\ell m}$ as
\begin{align}
	[\kappa^n]_{\ell m} &\equiv\int d\oh\, \kappa^{n}(\oh)Y_{\ell m}^*(\oh)\nonumber\\
	&=\sum_{\ell_1 m_1}\cdots\sum_{\ell_n m_n}\kappa_{\ell_1m_1}\cdots\kappa_{\ell_n m_n}\int d\hat\Omega\, Y_{\ell_1m_1}(\oh)\cdots Y_{\ell_n m_n}(\oh)Y_{\ell m}^*(\oh).
\end{align}
The integral involving spherical harmonics above can be expressed in terms of (products and sums of) the Wigner 3-j symbols. 
We can then define the following two {\it kurt-spectra}, which we will denote
as $K_{\ell}^{22}$ and ${K}_{\ell}^{31}$~\cite{myTri1, myTri2}:
\begin{align}
	{K}_\ell^{22} &\equiv\frac{1}{2\ell+1}\sum_m \langle[\kappa^2]_{\ell m}^{\phantom{*}}[\kappa^2]_{\ell m}^*\rangle=\sum_{\ell_1\cdots\ell_4} \frac{g^{\ell_1\ell_2\ell}g^{\ell_3\ell_4\ell}}{(2\ell+1)^2}T^{\ell_1\ell_2}_{\ell_3\ell_4}(\ell)\, , \label{C22} \\ 	
	{K}_\ell^{31} &\equiv\frac{1}{2\ell+1}\sum_m \Re\langle[\kappa^3]_{\ell m}^{\phantom{*}}\kappa_{\ell m}^*\rangle=\sum_{\ell_1\ell_2\ell_3 L} \frac{g^{\ell_1\ell_2L}g^{\ell_3\ell L}}{(2L+1)(2\ell+1)}T^{\ell_1\ell_2}_{\ell_3\ell }(L)\, ,\label{C31}
\end{align}
where we have expressed the results in terms of the weak lensing angular trispectrum $T^{\ell_1\ell_2}_{\ell_3\ell_4}(L)$ and $\Re$ denotes the real part.\footnote{
The relations between the kurt-spectra and the trispectrum can be derived by using the inversion formula for the angular trispectrum
\begin{align}
	T^{\ell_1\ell_2}_{\ell_3\ell_4}(L) =(2L+1)\sum_{M,m_i}\begin{pmatrix}
		\ell_1 & \ell_2 & L \\ m_1 & m_2 & M
	\end{pmatrix}\begin{pmatrix}
		\ell_3 & \ell_4 & L \\ m_3 & m_4 & -M
	\end{pmatrix}\langle \kappa_{\ell_1m_1}\cdots\kappa_{\ell_4m_4}\rangle\, .
\end{align}
}
These are natural generalizations of the skew-spectrum $\langle [\kappa^{2}]_{\ell m}^{\phantom{*}} \kappa_{\ell m}^* \rangle$ statistics studied in \cite{skew} to fourth order.\footnote{For the skew-spectra applied to galaxy statistics, see also~\cite{Schmittfull:2014tca,MoradinezhadDizgah:2019xun, Schmittfull:2020hoi}.}  

The Gaussian (disconnected) contribution to the trispectrum, which we denote by $G^{\ell_1\ell_2}_{\ell_3\ell_4}(L)$,
depends only on the angular power spectrum $C_{\ell}= \langle \kp_{\ell m}^{\phantom{*}}\kp^{*}_{\ell m}\rangle$, and is given by the expression~\cite{Hu_Trispectrum}
\begin{align}
G^{\ell_1\ell_2}_{\ell_3\ell_4}(L) & = (-1)^{\ell_1+\ell_3} \sqrt {(2\ell_1+1)(2\ell_3+1)}
    C_{\ell_1}C_{\ell_3}\delta_{L0}\delta_{\ell_1\ell_2}\delta_{\ell_3\ell_4} \nn
& \quad + (2L+1)C_{\ell_1}C_{\ell_2}\big [ (-1)^{\ell_2+\ell_3+L} \delta_{\ell_1\ell_3}\delta_{\ell_2\ell_4} + \delta_{\ell_1\ell_4}\delta_{\ell_2\ell_3}\big ]\, ,\label{eq:Gauss}
\end{align}
where $\delta_{\ell_a\ell_b}$ denotes the Kronecker delta. The corresponding Gaussian contribution to the kurt-spectra are given by
\begin{align}
G_{\ell}^{22} &= \sum_{\ell_1\ell_2\ell_3\ell_4} \frac{g^{\ell_1\ell_2\ell}g^{\ell_3\ell_4 \ell}}{(2\ell+1)^2} G^{\ell_3\ell_4}_{\ell_1\ell_2}(\ell)  \, , \label{eq:defG22}\\
G_{\ell}^{31} &= \sum_{\ell_1\ell_2\ell_3L} \frac{g^{\ell_1\ell_2\ell}g^{\ell_3L\ell}}{(2L+1)(2\ell+1)} G^{\ell_3\ell}_{\ell_1\ell_2}(L) \, . \label{eq:defG31}
\end{align}
These need to be subtracted from the total kurt-spectra. Substituting the Gaussian trispectrum \eqref{eq:Gauss}, we get
\ben
    && G^{22}_{\ell} = {1\over 4\pi} \sum_{\ell_1\ell_2} (2\ell_1+1)(2\ell_2+1)
    C_{\ell_1} C_{\ell_2}
    \left[ \delta_{\ell 0} + 2
      \begin{pmatrix}
     \ell_1 & \ell_2 & \ell \\
     0 & 0 & 0
      \end{pmatrix}^2
      \right], \label{eq:G22_theory}\\
    && G^{31}_{\ell} = 3C_{\ell}\times {1\over 4\pi }\,
    \sum_{\ell^{\prime}} (2\ell^{\prime} + 1) C_{\ell^{\prime}}\, .
    \label{eq:G31_theory}
    \een
These two spectra are related to the real-space kurtosis by
\ben
	\frac{1}{4\pi}\int d\hat n\,\langle\kappa^4(\hat n)\rangle = {3\over 4\pi}\sum_{\ell} (2\ell+1) G^{22}_{\ell} = {3\over 4\pi}\sum_{\ell} (2\ell+1) G^{31}_{\ell} = 3\sigma^4\, ,
\een
where $\sigma^2\equiv \frac{1}{4\pi}\int d\hat n\,\langle\kappa^2(\hat n)\rangle= \frac{1}{4\pi}\sum_\ell (2\ell+1)C_\ell$ is the angle-averaged variance.  To include the contribution from the noise, $C_{\ell}$ in \eqref{eq:Gauss} should be
  replaced by $C_{\ell} + N_{\ell}$. For Gaussian noise, the noise power spectrum is independent of $\ell$: $N_{\ell} = 4\pi\sigma^2/N_{\rm pix}$.

In addition to the kurt-spectra at fourth order, we have also computed the fifth-, sixth- and seventh-order
spectra from numerical simulations. At each order, there are more than one spectrum; 
for example, there are two fifth-order spectra defined as follows\footnote{
While the harmonic mode decomposition and a related power spectral analysis is ideal for a higher sky coverage, for ongoing surveys with a small fraction of sky coverage it is often easier to work in the real-space domain to avoid complications related to irregular mask or survey geometry.
The higher-order correlation functions corresponding to these high-order spectra are obtained by the usual Legendre transform as $C^{pq}(\theta) = \frac{1}{4\pi} \sum_{\ell} C^{pq}_\ell (2\ell+1) P_{\ell}(\cos\theta)$.
}
\ben
&& \Re\langle [\kappa^4]_{\ell m}^{\phantom{*}} \kappa^*_{\ell' m'}\rangle = C^{41}_{\ell}\delta_{\ell\ell^{\prime}}\delta_{mm'}\, ,\quad
\Re\langle [\kappa^3]_{\ell m}^{\phantom{*}} [\kappa^2]^*_{\ell' m'}\rangle = C^{32}_{\ell}\delta_{\ell\ell^{\prime}}\delta_{mm'}\, .
\label{eq:fifth}
\een
The triplets of sixth-order spectra  $C^{51}_{\ell}$,  $C^{42}_{\ell}$, $C^{33}_{\ell}$ and seventh-order spectra  $C^{61}_{\ell}, C^{52}_{\ell}$, $C^{43}_{\ell}$ are defined analogously.
Note that, unlike even-order spectra, there is no Gaussian contribution at odd orders. The addition of noise,
typically assumed to be Gaussian, increases the scatter at odd orders, while for even orders it affects
the mean of the estimator through its contribution to the disconnected components. 

\section{Comparison with Ray-Tracing Simulations}
\label{sec:disc}

Having described the calculation of the matter trispectrum and the weak lensing kurt-spectra, we now compare these theoretical signals to simulations. 
We first describe the details of the $N$-body simulations used in~\S\ref{sec:simu} and then discuss the results in \S\ref{sec:res}.

\subsection{Simulation Specifications}
\label{sec:simu}

We use the publicly available all-sky weak lensing maps generated by
\cite{Ryuchi}\footnote{\href{http://cosmo.phys.hirosaki-u.ac.jp/takahasi/allsky\_raytracing/}{\tt http://cosmo.phys.hirosaki-u.ac.jp/takahasi/allsky\_raytracing/}} using a ray-tracing scheme through $N$-body simulations.
The underlying $N$-body simulations follow the gravitational clustering of $2048^3$ particles.
Multiple lens planes were used to generate the lensing convergence $\kappa$ 
and the corresponding shear $\gamma$ maps. 
To generate the maps in these simulations, the source redshifts used were in the range $z_s\in[0.05,5.30]$ at a redshift-interval of $\Delta z_s = 0.05$.
In this study, we have used the maps with  source-redshifts $z_s=0.5$, $1$, and $2$, using the following fiducial cosmological parameters: the dimensionless Hubble parameter $h=0.7$, the dark matter density $\Omega_{cdm} = 0.233$, the baryon density $\Omega_{b} = 0.046$,
the matter density $\Omega_m = \Omega_{cdm}+\Omega_b$, the amplitude of matter fluctuations on 8$h^{-1}$Mpc scales $\sigma_8=0.82$, the scalar spectral index $n_s=0.97$, and a flat universe. In a previous study~\cite{myskew12}, inclusion of post-Born terms in lensing statistics were studied at the level of the bispectrum.
Although post-Born corrections play a significant role at higher redshift, e.g.~in the case of CMB lensing, it was found that such corrections play a negligible role at the low-source redshifts that we study in this work.

The lensing convergence maps were generated using an equal area pixelization scheme
in {\tt HEALPix}\footnote{\href{https://healpix.jpl.nasa.gov/}{\tt https://healpix.jpl.nasa.gov/}} format \cite{Gorski}.
In this pixelization scheme, the number of pixels scales as $N_{\rm pix} = 12 N^2_{\rm side}$. The resolution parameter $N_{\rm side}$ can take values $N_{\rm side} = 2^{m}$ with $m=1,2,\cdots$.
The maps used in this study are generated at $N_{\rm side}=4096$ and were cross-checked against higher-resolution maps constructed at $N_{\rm side}=8192$, $16384$ for consistency, up to $\ell_{\rm max}= 2\times 10^3$. Many
additional tests were also performed using the $E/B$ decomposition of the shear maps
for the construction of $\kappa$ maps \cite{Ryuchi}.
After this validation procedure, we have degraded these maps to $N_{\rm side}= 1024$ and analyzed them for harmonic modes satisfying $\ell \le 2N_{\rm side}$.

\begin{figure}
\begin{center}
  \includegraphics[width=0.95\textwidth]{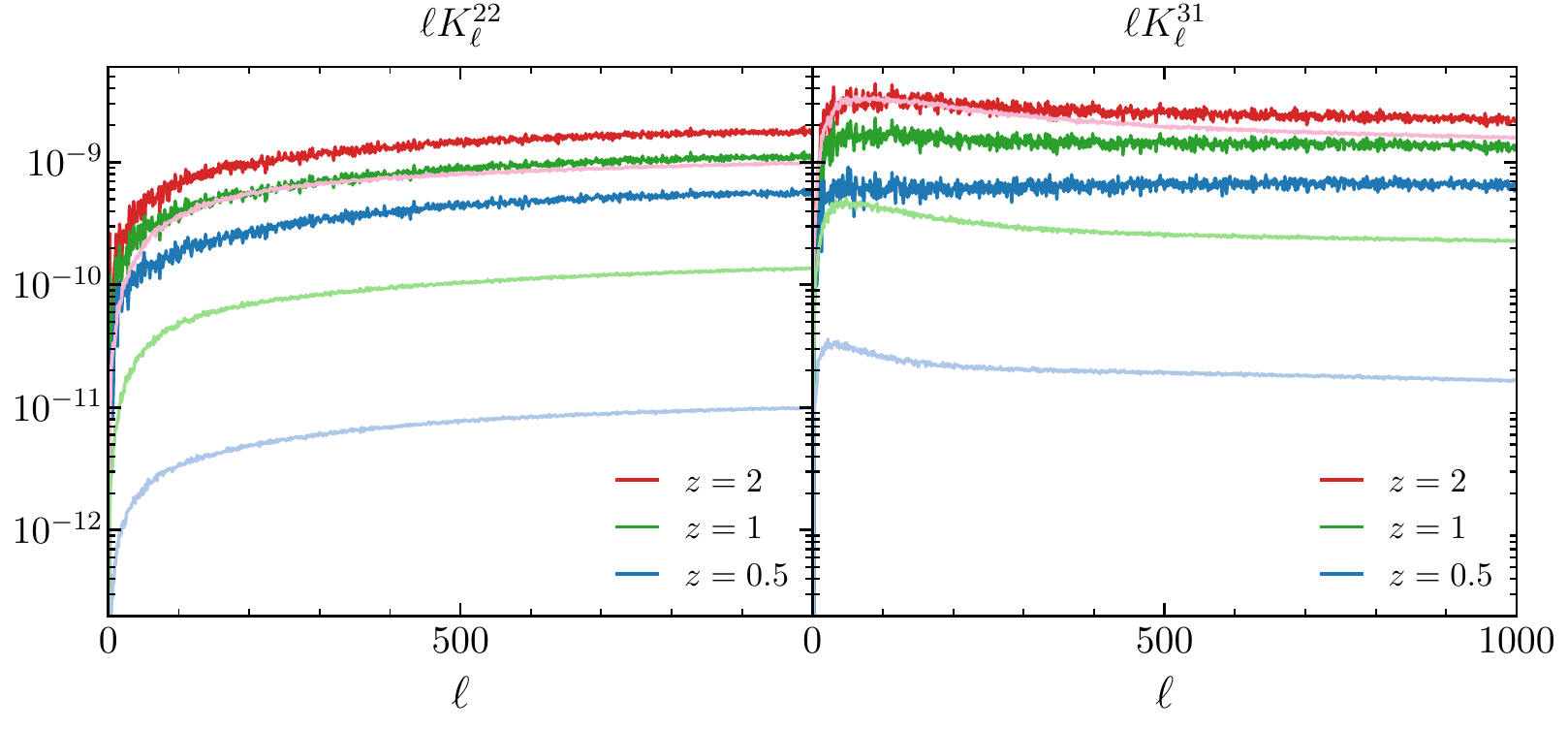}
  \end{center}
  \vspace{-0.25cm}
  \caption{Kurt-spectra $K^{31}_{\ell}$
    and $K^{22}_{\ell}$ as defined in \eqref{C22}
    and \eqref{C31}, without any beam smoothing.
    In each panel, the (pale) red, green, and blue curves show the (Gaussian) kurt-spectra for source redshifts $z_s=0.5$, $1$, and $2$, respectively. One single realization with $N_{\rm side}=1024$ was used to generate the total kurt-spectra, while the Gaussian parts are generated using ten realizations of Gaussian maps from the theoretical power spectrum.}
  \label{fig:kurt}
\end{figure} 

\subsection{Results and Discussion}
\label{sec:res}

\subsubsection{Shapes of Kurt-Spectra}

In Figure \ref{fig:kurt}, we show the redshift dependence and the shapes
  of the two kurt-spectra defined in \eqref{C22}
  and \eqref{C31} from a single realization, without an observational mask or noise.
  The various lines, from bottom to top in each panels,
  present the results for source redshifts $z_s=0.5$, $1$, and~$2$, respectively. We see that the amplitude of $K^{31}_\ell$ is typically higher than that of $K^{22}_\ell$ at low $\ell$. This is because a larger number of non-vanishing trispectrum configurations contribute to the former at a given $\ell\ll\ell_{\rm max}$.
  
  In addition to the total kurt-spectra, we have also generated ten Gaussian realizations to estimate the contribution to the kurt-spectra from
  the disconnected parts of the trispectrum, \eqref{eq:defG31} and \eqref{eq:defG22}, which are shown in pale-colored lines. 
  For generating these realizations, we have used the same power spectra as the original numerical simulations.
  As expected, the Gaussian contributions are subdominant at low redshifts where $\kappa$ traces the highly non-Gaussian underlying density distribution. 
  These simulated Gaussian kurt-spectra were found to match the theoretical signals accurately.

\subsubsection{Observational Mask and Noise}

\begin{figure}
\begin{center}
  \includegraphics[width=0.9\textwidth]{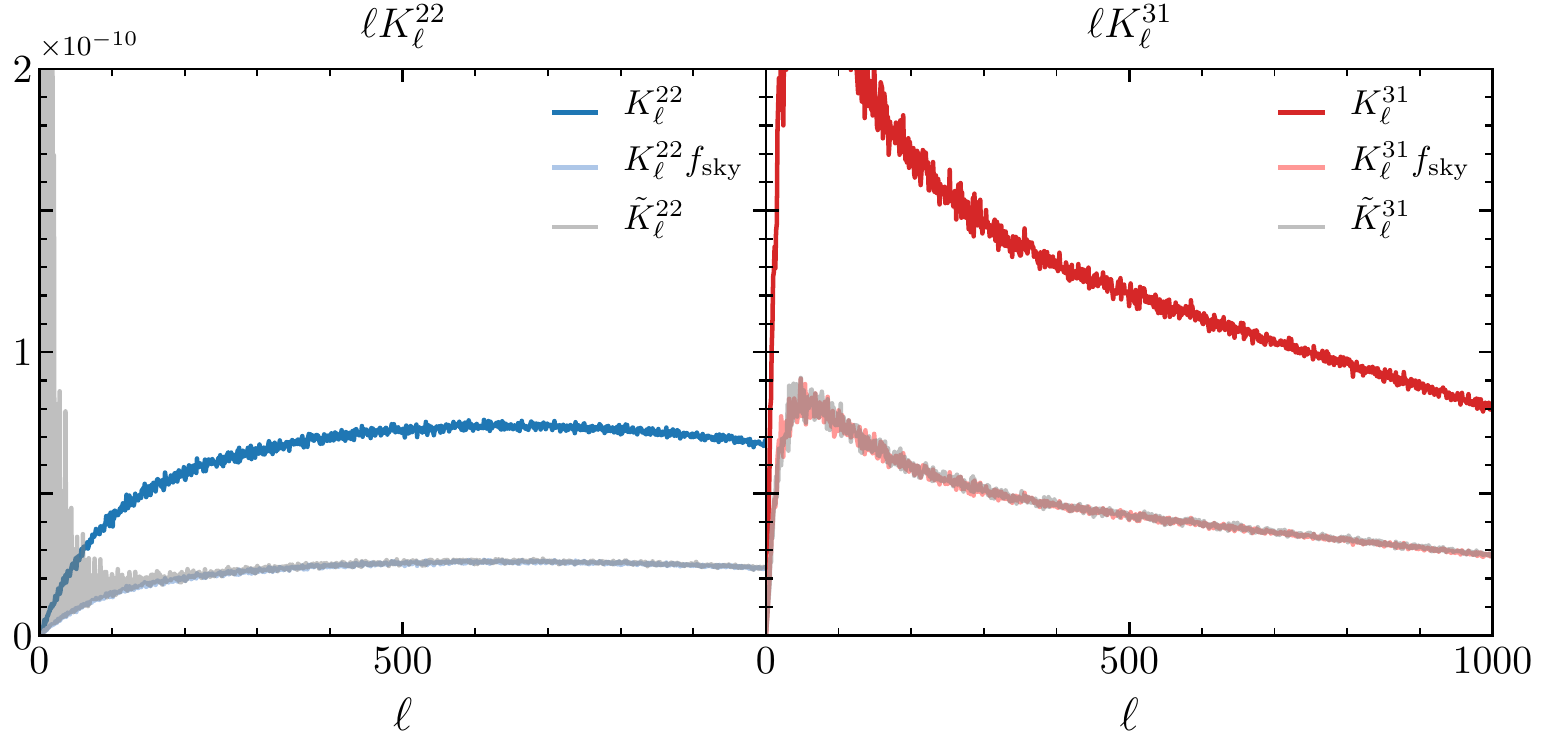}
  \end{center}
  \vspace{-0.25cm}
  \caption{Kurt-spectra at $z_s=1$ after applying a {\em Euclid}-type mask, with no noise inclusion.
    The smoothing angular scale is fixed at $\theta_s=5'$. The upper curve shows $K^{pq}_\ell$ computed
    using a single realization  without the observational mask. The two nearly-identical curves at the bottom show the corresponding masked spectra, ${\tilde{K}}^{pq}_\ell$, and
    the unmasked spectra multiplied with the fraction of sky coverage, $K^{pq}_\ell f_{\rm sky}$. 
   }
  \label{fig:mask} 
\end{figure} 

   Observational masks introduce mode couplings that need
    to be corrected before studying the gravity-induced mode coupling. 
    An efficient approach was introduced in~\cite{hivon}
    to study the ordinary angular power spectrum, commonly known as the
    pseudo-$C_{\ell}$ or PCL technique. Using the PCL approach,
    an unbiased estimator for the $(p+q)-$th order power spectrum, $\hat C_{\ell}^{pq}$, can be
    expressed as
    \ben
    \hat C^{pq}_\ell =
    \sum_{\ell'}M^{-1}_{\ell\ell^\prime}( \tilde C^{pq}_{\ell^{\prime}}- \tilde G^{pq}_{\ell^{\prime}} )\, ,
    \een
    where $\tilde C^{pq}_{\ell}$ represents the total spectrum
    estimated from a noisy map in the presence of mask 
    \ben
    \tilde C^{pq}_\ell =
    {1\over 2\ell+1}\sum_{m} {{[\tilde\kappa}^p]_{\ell m}}{{[\tilde\kappa}^q]_{\ell m}}\, ,
    \een
    with  $\tilde\kappa$ the 
    masked $\kappa$ map. The Gaussian component
    of the spectrum is denoted above as
    $\tilde G^{pq}_{\ell}$, which is computed using Monte Carlo realizations of Gaussian maps in the presence of the same mask and noise.
     The matrix $M_{\ell\ell'}$ that encodes the mode-coupling information induced by the mask takes the form
\begin{equation}
M_{\ell\ell'} = (2\ell'+1)\sum_{\ell''}
\frac{2\ell'' +1}{4\pi} |w_{\ell''}|^2 \begin{pmatrix}
    \ell & \ell' & \ell'' \\
     0 & 0 & 0
\end{pmatrix}^2\, ,
\end{equation}
where $w_{\ell''}$ represents the angular power spectrum of the survey mask. In the high-$\ell$ regime, the
    coupling matrix simplifies as 
    $M_{\ell\ell^\prime} \approx f_{\rm sky}\delta_{\ell\ell^\prime}$ with $f_{\rm sky}$ the fraction of sky coverage.
    For the spectra of order higher than four, terms involving lower-order spectra will contribute
      and generation of Gaussian maps may not be enough to subtract the disconnected contributions.

Using this technique, we study the effect of a {\em Euclid}-type mask in the estimation of the kurt-spectra. (For reference, the mask we have used has $f_{\rm sky}\approx 0.35$ and is described in \cite{skew}.) The results are shown in Figure \ref{fig:mask}, where the left and right panels show the kurt-spectra ${K}^{22}_\ell$ and ${K}^{31}_\ell$, respectively. The smoothing angular scale is fixed at $\theta_s=5'$ in both cases.  
    In each panel, the upper curve shows $K^{pq}_\ell$ computed
    from a single realization without any noise added. The two nearly-identical
    curves at the bottom show the corresponding masked ${\tilde{K}}^{pq}_\ell$ and
    the rescaled unmasked spectra~$f_{\rm sky}K^{pq}_\ell$. 
    The same scaling with $f_{\rm sky}$ can be applied to the Gaussian contribution
    and the noise; hence these contributions can simply be subtracted to construct an unbiased estimator. We used a sharp mask without any apodization. The large-scale features of the mask then appear as fluctuations in the convolved spectra, which survive the auto-spectrum $\tilde K^{22}_\ell$ but not in the cross-spectrum $\tilde K^{31}_\ell$.

 The estimator we have introduced here is a sub-optimal estimator.\footnote{The flat-sky equivalent of PCLs used here was developed in~\cite{chiaki}.} This is sufficient for all-sky surveys where the signal-to-noise is very high. A nearly-optimal estimator which is also unbiased was considered for PNG in the CMB
in~\cite{myTri1, myTri2}. This method depends on applying weights that depend on the target trispectrum and is computationally more expensive. An optimal method was also presented in \cite{myTri1, myTri2}, which involves inverse covariance weighting.
Such estimators are optimal only in the limit of small non-Gaussianity (e.g.~PNG).
However, such an approach is neither realistic nor necessary for secondary non-Gaussianity where the non-Gaussian signal is quite strong.   
Optimized versions of the kurt-spectra have also been considered in \cite{myTri1} for PNG, though they cannot be estimated using a PCL estimator. In the presence of a mask, the linear correction terms require a more elaborate Monte Carlo computation involving Gaussian random realizations. Optimization of our estimator to the gravity-induced secondary non-Gaussianity will be presented elsewhere.  

    \begin{figure}[t!]
  \begin{center}
      \includegraphics[scale=0.85]{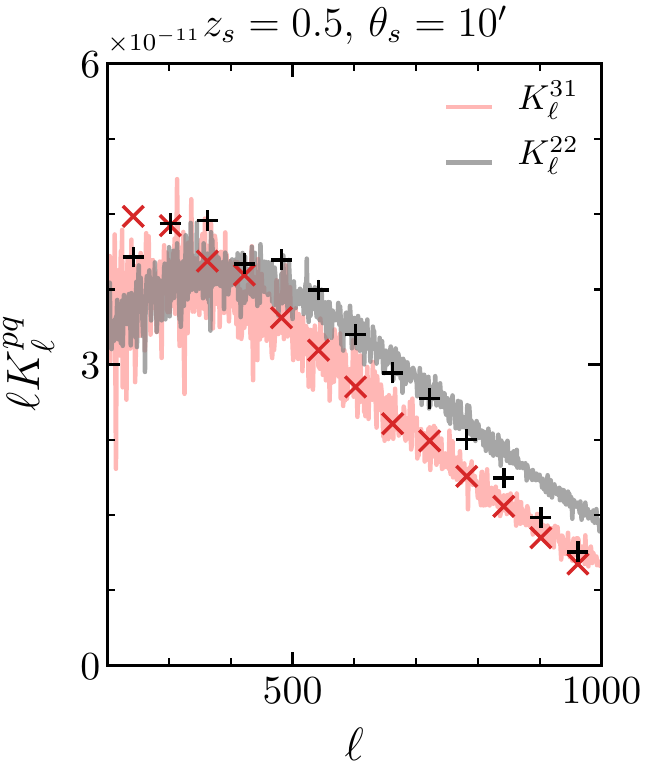}
\hspace{-.4cm}      \includegraphics[scale=0.85]{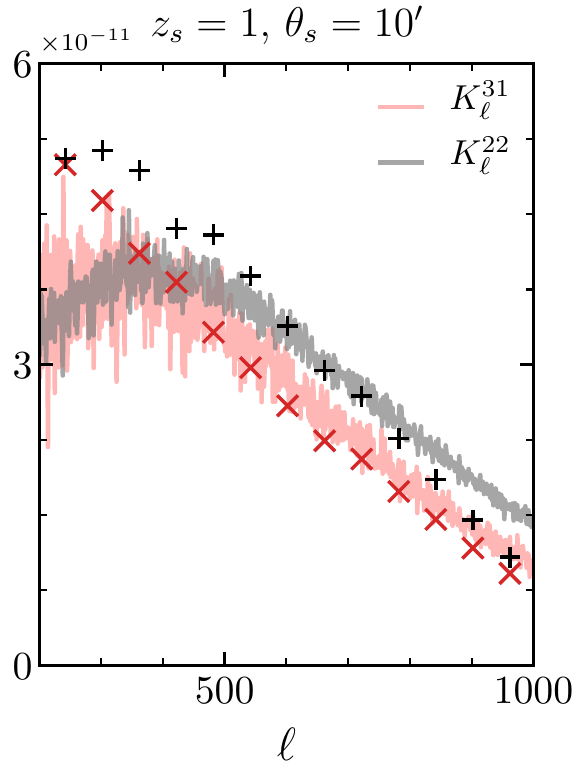}
\hspace{-.4cm}      \includegraphics[scale=0.85]{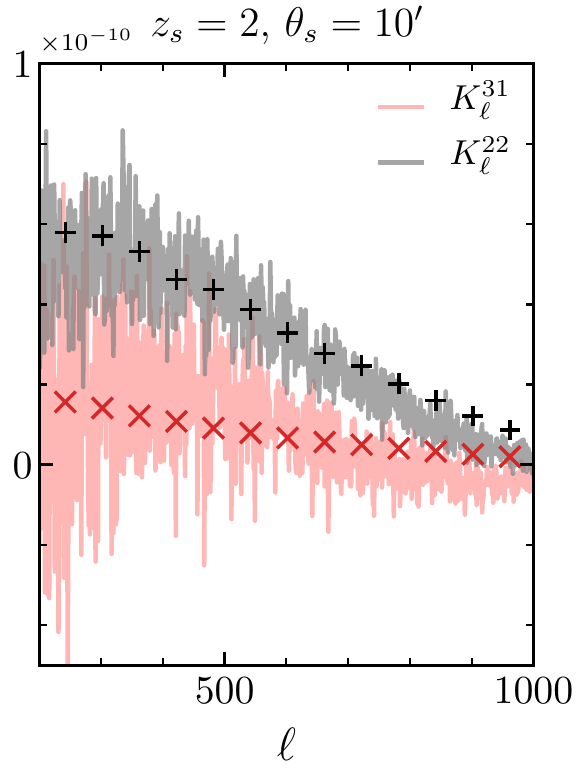}
\end{center}
  \vspace{-0.25cm}
  \caption{Comparison of theoretical and numerical results for the kurt-spectra with $\theta_s=10'$. The discrete points
    represent theoretical predictions using the hierarchical ansatz. The simulation results are an average of
    ten realizations (see the text for more details).
  }\label{fig:comp1}
\end{figure} 


\subsubsection{Comparison with Theory}

\begin{figure}[t!]
  \begin{center}
      \includegraphics[scale=0.85]{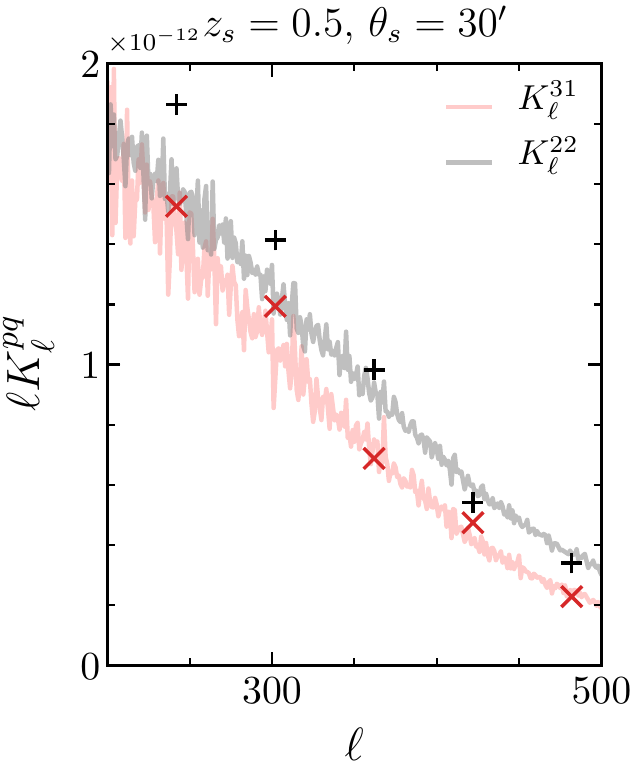}
\hspace{-.4cm}      \includegraphics[scale=0.85]{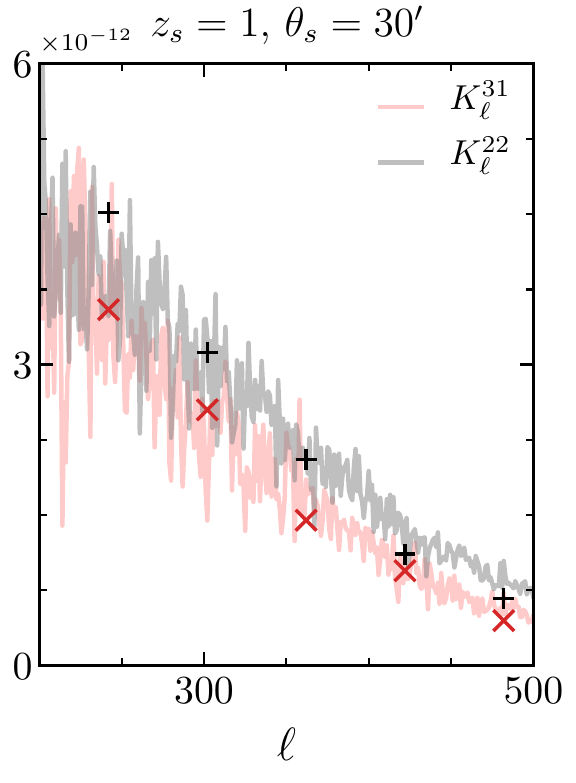}
\hspace{-.4cm}      \includegraphics[scale=0.85]{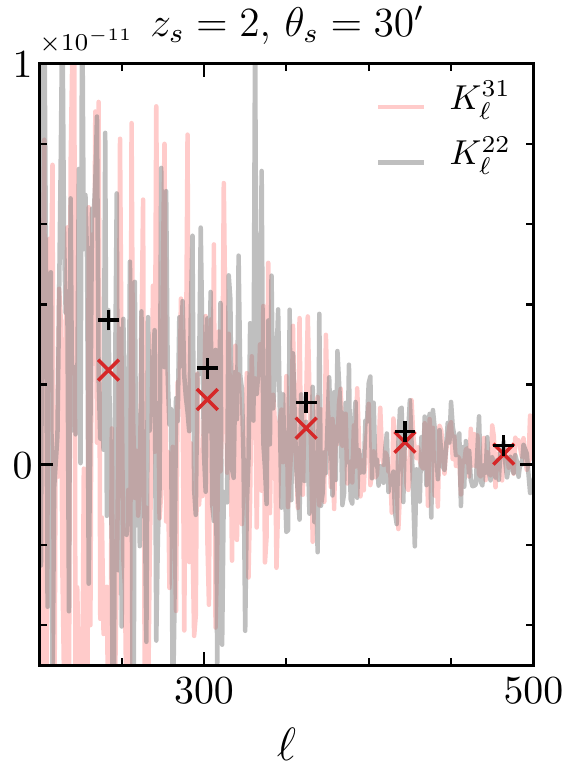}
\end{center}
  \vspace{-0.25cm}
  \caption{Comparison of theoretical and numerical results for the kurt-spectra with $\theta_s=30'$. The discrete points
    represent theoretical predictions using the hierarchical ansatz. The simulation results show average of
    ten realisations (see text for more details).}\label{fig:comp2}
\end{figure}

  In Figures \ref{fig:comp1} and~\ref{fig:comp2}, we present the results of our comparison of a theoretical model against numerical
  simulations for source redshifts $z_s=0.5$, 1, and 2. We have used two different smoothing angular scales $\theta_s=10'$ and 
  $\theta_s=30'$, for which it was sufficient to use the maps with
  $N_{\rm side} =512$ and $\ell_{\rm max}=1024$. Two curves are shown in each panel: the red and gray curves represent ${K}^{31}_{\ell}$  and ${K}^{22}_{\ell}$, respectively. For the purpose of comparing against purely theoretical signals from the connected trispectrum, we have subtracted the Gaussian contributions and no noise and mask were used. The discrete points represent the theoretical results using the HA given in \eqref{THA} with $R_a=R_b$ as a free parameter to fit the data. As can be seen from the figures, the kurt-spectra resulting from the HA agree reasonably well with the simulation results in the nonlinear regime, while it has an upward trend and starts to display a large deviation towards low multipoles.\footnote{In general, low multipoles are affected by the finite size of the survey volume, which is more pronounced in higher-order statistics. The ray-tracing simulations inherit the finite volume corrections from the $N$-body simulations used to generate the lensing maps. For this reason, we have mostly concentrated on harmonics $\ell >100$, as in the previous work~\cite{skew, shape}.}
  
We have also done a comparison with the tree-level SPT trispectrum used as the input, and found that they deviate significantly from the simulation results even at low $\ell$, with or without using the Limber approximation. This failure can be attributed to the fact that the tree-level approximation typically remains valid up to the scale $k_\star\approx 0.1\, h\hskip 1pt \text{Mpc}^{-1}$ in Fourier space, with the corresponding nonlinear multipoles $\ell_\star\approx 90$, 150, and 240 for $z_s=0.5$, 1, and 2, respectively. The kurt-spectra therefore involve summing over a large number of nonlinear modes for the smoothing scales that we have considered in this work. For example, even the largest smoothing scale $\theta_s=30'$ that we used only reduces the amplitude at $\ell_\star=240$ by 30\%, which is not sufficient to suppress the contribution from nonlinear modes.

The HA and SPT both generate similar correlation structure but with different hierarchical amplitudes
  $R_a$ and $R_b$ \cite{bernardeaureview}. This explains the fact that in Figure \ref{fig:comp1} the theoretical predictions
better match numerical simulations for source redshift $z_s=0.5$ (highly nonlinear regime)
and $z_s=2.0$ (quasi-linear regime). However, in the intermediate regime, the form of the correlation hierarchy
is more complicated and remains poorly understood.
This is reflected in the middle panel of Figure {\ref{fig:comp1}} for $z_s=1.0$.
Indeed, the line-of-sight integral mixes various modes.

There are clear deviations in the high-$\ell$ regime especially for $z_s=2.0$. The HA can be extended to consider
  $\epsilon\ne0$ in Eq. (\ref{THA}).  However, our aim in this article is not to provide a detailed phenomenological
  fitting function, but rather to introduce the higher-order spectra in the analysis of weak lensing maps.

Extending the calculation to one loop order would allow us to include modes up to $k_\star\approx 0.3\, h\hskip 1pt \text{Mpc}^{-1}$ and would significantly extend the range of validity of perturbation theory in harmonic space (e.g.~up to $\ell_\star\approx 720$ at $z_s=2$). It would thus be interesting to compute the kurt-spectra including the contribution from the one-loop matter trispectrum in the EFT framework~\cite{Bertolini:2015fya, Bertolini:2016bmt, Steele:2021lnz} and compare its validity against simulations. We leave this for future work. 

It is instructive to compare the problem at hand with the computation of the non-Gaussian covariance of the angular power spectrum, which takes the form~\cite{Hu_Trispectrum}
\begin{align}
    C_{\ell\ell'} = \frac{(-1)^{\ell+\ell'}}{\sqrt{(2\ell+1)(2\ell'+1)}}T^{\ell\ell}_{\ell'\ell'}(0)-C_\ell C_{\ell'}\, .
\end{align}
Notice that, unlike the kurt-spectra, this does not involve summing over nonlinear modes,\footnote{Note that $T^{\ell_1\ell_2}_{\ell_3\ell_4}(L)$ is the trispectrum defined with a specific channel decomposition, so it in principle includes a sum over nonlinear modes from other channels, see \eqref{Tdecomp}. As explained earlier, these contributions are, however, usually highly suppressed in the limit $L\to 0$ and so can be neglected.} and it is known that the tree-level approximation provides a good approximation at low multipoles~\cite{Scoccimarro:1999kp, LeeDvorkin}. Moreover, the non-Gaussian covariance involves taking the infrared multipole $L=0$, so it always receives contributions from large scales, making the HA inadequate in this case; see also~\cite{Scoccimarro:1999kp}.

\subsubsection{Beyond Fourth Order}
\begin{figure}
   \begin{center}
      \includegraphics[scale=0.85]{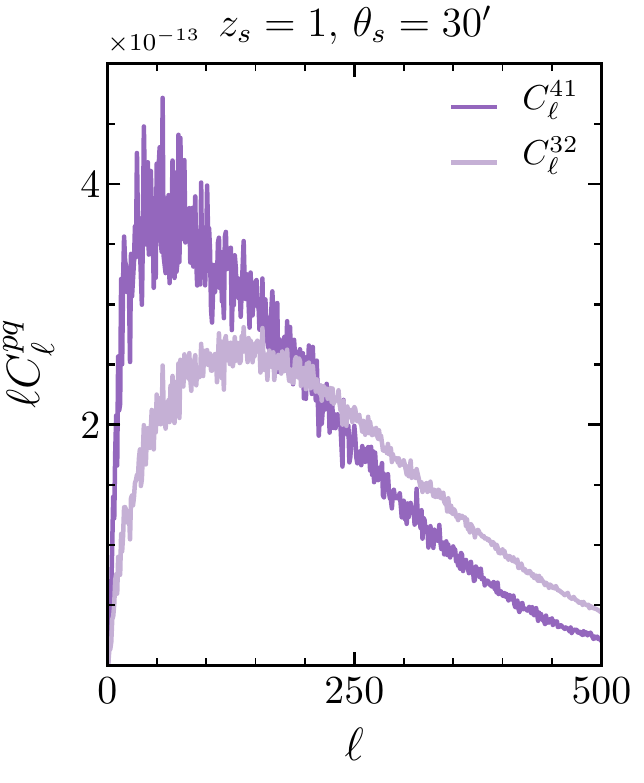}
\hspace{-.4cm}      \includegraphics[scale=0.85]{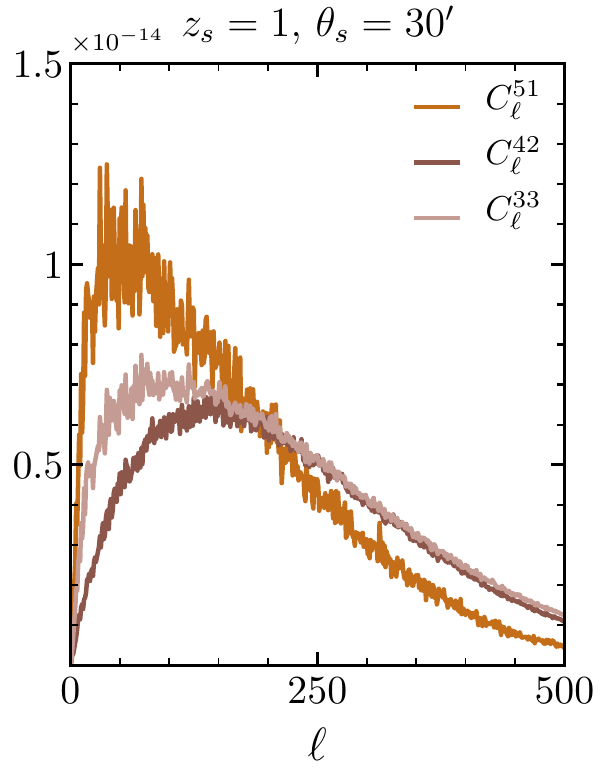}
\hspace{-.4cm}      \includegraphics[scale=0.85]{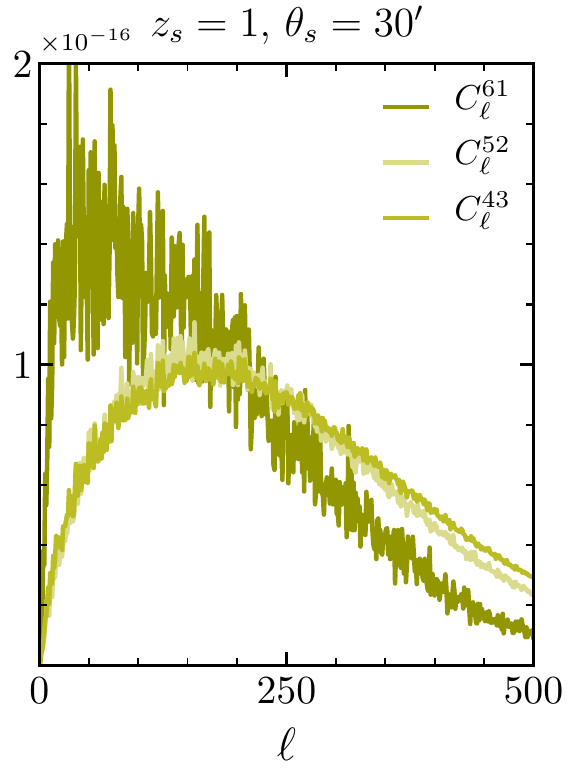}
\end{center}
  \vspace{-0.25cm}
  \caption{Higher-order spectra of weak lensing convergence at $z_s=1$ with $\theta_s=30'$. The different panels from left to right
    show the fifth-, sixth-, and seventh-order spectra.
 }
  \label{fig:beyond}
\end{figure} 

  We have also simulated spectra beyond fourth order, as shown in Figure \ref{fig:beyond}.
  From left to right the different panels show 
  the spectra at fifth, sixth, and seventh orders, with the smoothing scale $\theta_s=30'$ and the source redshift $z_s=1$. The original maps created at a {\tt HEALPix} resolution of $N_{\rm side}=4096$ were degraded
    to $N_{\rm side}=1024$ before the analysis, and a total of ten realizations were used. All-sky maps were considered and no noise was added to the maps. The Gaussian contributions to the sixth-order spectra are also shown, while there are no Gaussian parts at odd orders.
  
  We do not have an accurate analytical model to compute the scatter in the spectra.
  However, we know that the higher the power of the spectra, the more scatter it will present. This is because higher-order spectra probe the tails of the $\kappa$-distribution, so they are more susceptible to the presence (or absence) of rare high (or low) $\kappa$ values.
  As a consequence, $C_\ell^{41}$ will be noisier as it
  contains fourth power compared to, say, $C_\ell^{32}$,
  which cross-correlates fields with lower powers. Currently, there are
  no well-established estimator for higher-order non-Gaussianities beyond fourth order. Nevertheless, from the figures it is clear that surveys such as {\it Euclid} will be able to probe these non-Gaussian spectra beyond fourth order. 
  Utilizing these higher-order spectra can help tighten cosmological parameter constraints.
  
The stage-IV weak lensing surveys will be in the {signal-dominated} regime \cite{Taylor:2019mgj}. Estimating the signal-to-noise involves computing the error covariance matrix of the PCL estimator that we have considered. This is nontrivial even in the context of the ordinary power spectrum in the signal dominated regime. Theoretical computation of the covariance matrix requires an analytical modeling of even high-order spectra, e.g.~eight-order correlations for the covariance of the kurt-spectra estimators. 
This is currently not possible in a reliable manner using the HA or extensions of halo models. One possible option is to use simulations to model the higher-order correlations. Indeed, to get a reliable estimate of the off-diagonal terms in the covariance matrix, an increased number of simulations will be required. This remains an active area of research.

\subsubsection{Low-$\ell$ modes and Finite Volume Corrections}

In the low-$\ell$ regime our model over predicts simulation results for both $K^{22}_\ell$ and $K_{\ell}^{31}$.
  This is related to the fact that in this regime
  at least one leg of the trispectrum is in the
  perturbative regime. The magnitude associated with the perturbative
  trispectrum for a given configuration is expected to be lower than its hierarchical counterparts.
  
 In addition, it is worth mentioning that the effect of the finite volume of
   the simulation is known to play an important role for the
  determination of one-point statistics (see, e.g., \cite{Szapudi_Colombi}).
  Previous studies focused on one-point moments, but the spectra we are
  constructing are two-point statistics. At the moment, to the best of our knowledge,
  there is no prescription
  to correct the bias due to such finite volume corrections for two-point statistics.

\section{Conclusions and Future Prospects}
\label{sec:conclusions}

Most studies of weak lensing non-Gaussianity focus on the leading-order non-Gaussianity, namely the bispectrum. 
In this paper, we have extended these works to fourth-order statistics by introducing two new fourth-order spectra called kurt-spectra that generalize the concept of kurtosis---the fourth-order cumulant---in harmonic space. 
We have used pseudo-$C_{\ell}$-based estimators that can estimate 
these kurt-spectra from realistic weak-lensing maps that involve an observational mask and noise. We have shown how the Gaussian
components of these spectra can be subtracted using Monte-Carlo realizations of Gaussian maps or the theoretical expectation. One of the main outcomes of our study is the
fact that the kurt-spectra, as well as their higher-point generalizations, can be reliably
extracted from (nearly-)all-sky weak lensing surveys.

Additionally, we have introduced a framework
to compute these statistics theoretically. However, we 
found that obtaining an exact matching with the simulation results is not straightforward for two different reasons.
At the level of the bispectrum, there currently exist halo-model-based numerical fitting functions that can be used to accurately
predict the skew-spectrum. On the other hand, we do not have a such numerical fitting function for the trispectrum or beyond. 
In circumventing this problem, we have outlined two different (analytical) approaches in this paper based on the SPT and the HA. The SPT is only valid for large smoothing scales, while HA is expected to be valid at much smaller length scales. The kurt-spectra involve a mode sum that mixes different scales, thus underlining the case for a fitting function
to reproduce the simulation results. 
An additional complexity is that, while computation of the skew-spectrum requires $O(\ell^2_{\rm max})$ evaluation of 
the bispectrum, the number of computation
for the kurt-spectra is $O(\ell^3_{\rm max})$, hence restricting the resolution of maps that can be analyzed. 

We have analyzed maps with $N_{\rm side}=512$ and $\ell_{\rm max}=1024$, and compared the simulation results with the theory predictions. With this choice, we found that the HA, despite its simplicity, can predict the general trends of the fourth-order spectra to a good accuracy in the nonlinear regime. At the same time, they show a pronounced departure from the numerical simulations at low multipole moments, due to the invalidity of the HA in this regime. We also found that the tree-level SPT trispectrum cannot be used to reliably predict the shapes of the kurt-spectra, which involve contributions from modes at $\ell\gg 100$. A better theoretical modeling is thus required to accurately compute the kurt-spectra, such as including the EFT trispectrum at one loop~\cite{Bertolini:2015fya, Bertolini:2016bmt, Steele:2021lnz} in the perturbative calculation. Another possibility is to use an emulator-based approach for cosmological statistics to avoid modeling of a fitting function~(see e.g.~\cite{SpurioMancini:2021ppk}).

The results presented here correspond to a single
source plane. In practice, the sources are distributed over a range of redshifts, which can be easily incorporated in our modeling. Future cosmological galaxy surveys, such as the Vera Rubin Observatory, will observe a very large number of galaxies. In the absence of spectroscopic data, their redshifts will have to be inferred from the photometric redshifts (photo-z). We leave incorporating the photo-z error in our modeling and study its implications in the future.

Along with the skew-spectrum, the kurt-spectra introduced in this paper will be useful in testing various mass-mapping techniques that are generally
employed. It is well known that the naive mass-mapping technique uses a flat-sky approximation known also as the Kaiser-Squires (KS) method~\cite{KS},  
which is an inversion of the forward model in the Fourier domain. 
This, however, does not take into account 
 noise or boundary effects. These are typically post-processed
 via convolutions that involve a large Gaussian smoothing kernel.
 This results in a heavy degradation of the quality of the non-Gaussian
 information content of high-resolution maps. 
 In addition, there are issues related to the fact that the decomposition of spin-fields into $E/B$ modes when performed on a bounded manifold is known to be degenerate. 
 It is thus commonly believed that the KS estimator can perform poorly in the presence of a nontrivial mask. 
 In recent years, a sparse hierarchical Bayesian formalism for all-sky  mass-mapping without making any assumptions or impositions of Gaussianity was developed (e.g. in~\cite{Price:2020mry}).
 The estimators developed here for higher-order
 statistics can be used to compare the reproducibility of the non-Gaussian 
 information in mass-mapping in the presence of a mask and noise.
 
We have studied a few representative trispectrum configurations. 
Different configurations of correlation functions are associated with
features in the large-scale structure, such as pancakes, filaments, clumps, voids, cosmic strings, as well as statistical anisotropy. Since the kurt-spectra reduce the entire shape information to one dimension, the study of various shapes of the trispectrum along with the kurt-spectra would thus yield a rich dividend. 
Based on the techniques developed in \cite{Namikawa:2018bju}, an estimator that can probe individual bispectrum shapes was proposed in \cite{shape}. We plan to generalize this estimator to explore the full shape of the trispectrum in future work.

\paragraph{Acknowledgments}
DM was supported by a grant from the Leverhulme Trust at MSSL when this project was initiated. HL and CD were partially supported by Department of Energy (DOE) grant DE-SC0020223. HL is supported by the Kavli Institute for Cosmological Physics through an endowment from the Kavli Foundation and its founder Fred Kavli. 
We would like to thank Peter Taylor for providing us his code to generate
the Euclid-type mask used in our study. 
We would also like to thank Ryuichi Takahashi for making his simulations
publicly available.

\newpage
\bibliographystyle{utphys}
\bibliography{weaklensing4pt}

\end{document}